# Ion beam modification of magnetic tunnel junctions

B. M. S. Teixeira[1,*], A. A. Timopheev[2], N. Caçoilo[1,3], L. Cuchet[2], J. Mondaud[2], J. R. Childress[2], S. Magalhães[4], E. Alves[4], N. A. Sobolev[1]

[1] i3N, Departamento de Física, Universidade de Aveiro, 3810-193, Aveiro, Portugal

[2] Crocus Technology, 3 avenue Doyen Louis Weil, BP1505 - 38025 GRENOBLE Cedex1, France

[3] Univ. Grenoble Alpes, CEA, CNRS, Grenoble INP, SPINTEC, F-38000 Grenoble, France

[4] IPFN, Instituto Superior Técnico, Universidade de Lisboa, 2695-066 Bobadela LRS, Portugal

* Corresponding author: bmsteixeira@gmail.com

## Abstract

The impact of 400 keV Ar$^+$ ion irradiation on the magnetic and electrical properties of in-plane magnetized magnetic tunnel junction (MTJ) stacks was investigated by ferromagnetic resonance, vibrating sample magnetometry and current-in-plane tunneling techniques. The irradiation-induced changes of the magnetic anisotropy, coupling energies and tunnel magnetoresistance (TMR) exhibited a correlated dependence on the ion fluence, which allowed us to distinguish between two irradiation regimes. In the low-fluence regime, $\Phi < 10^{14}$ cm$^{-2}$, the parameters required for having a functioning MTJ were preserved: the anisotropy of the FeCoB free layer (FL) was weakly modulated following a small decrease in the saturation magnetization $M_S$; the TMR decreased continuously; the interlayer exchange coupling (IEC) and the exchange bias (EB) decreased slightly. In the high-fluence regime, $\Phi > 10^{14}$ cm$^{-2}$, the MTJ was rendered inoperative: the modulation of the FL anisotropy was strong, caused by a strong decrease in $M_S$, ascribed to a high degree of interface intermixing between the FL and the Ta capping; the EB and IEC were also lost, likely due to intermixing of the layers composing the synthetic antiferromagnet; and the TMR vanished due to the irradiation-induced deterioration of the MgO barrier and MgO/FeCoB interfaces. We demonstrate that the layers surrounding the FL play a decisive role in determining the trend of the magnetic anisotropy evolution resulting from the irradiation, and that an ion-fluence window exists where such a modulation of magnetic anisotropy can occur, while not losing the TMR or the magnetic configuration of the MTJ.

**Introduction**

One advantage of the magnetic random-access memory (MRAM) compared to technologies that rely on electric charge for information storage (e.g. dynamic RAM) is its superior radiation hardness with respect to gamma rays and charged particles in the MeV range. That hardness makes MRAM promising for applications in extreme environments[1]. Still, magnetic tunnel junctions (MTJs) are not tolerant to all sorts of radiation[2–4]: indeed, ion-irradiation-induced modifications of MTJs have been observed, extending from soft errors (undesired but recoverable magnetization switching, provoked by localized heating[4]) to permanent changes in magnetic and electrical properties produced by structural modifications. The degree of such modifications is governed by the spatial profile of the total energy deposited in the materials, which is dependent on the ion mass and charge state, kinetic energy, fluence, target composition and target temperature. Various ion species (e.g. $He^+$, $Ga^+$ and $Ar^+$), with energies ranging from hundreds of eV to hundreds of MeV and fluences $\Phi$ between $10^{11}$ cm$^{-2}$ and $10^{17}$ cm$^{-2}$, have been used to purposefully modify properties of magnetic multilayers including MTJs.

The research regarding the tailoring of magnetism via ion irradiation published until 2004 was reviewed in Ref.[5]. Some noteworthy results on the control of magnetism by ion irradiation include the reorientation of the magnetization direction from out-of-plane to in-plane and also to oblique orientations in Pt/Co multilayers[6–8]; control of magnitude and direction of the exchange-bias field at ferromagnet/antiferromagnet interfaces[9–11]; changes of the Néel coupling via ion-beam smoothing of interfaces[12]; reduction of the annealing temperature required for crystallizing CoFeB in MgO-based MTJs[13]; decrease in critical current density for spin-orbit torque switching in Pt/Co/Ta[14]; improvement of the microwave emission linewidth of a spin-torque nano-oscillator[15], and tuning of the types[16] and of the velocity of propagation[17] of domain walls. Ion irradiation further enables lateral patterning which has been proposed for the definition of magnetoresistive sensors[5] as well as exploited for creating skyrmions[18,19] and for producing magnonic crystals for controlling the spin wave propagation[20–22].

In a previous work[23] we demonstrated that 400 keV $Ar^+$ ion irradiation can induce the easy-cone anisotropy in initially perpendicularly magnetized MgO/FeCoB/X/FeCoB/MgO stacks (X = Ta or W spacer). Such an easy-cone anisotropy is sought after as it can lead to faster and lower-energy spin-transfer torque (STT) switching, thanks to the intrinsic tilt in magnetization direction provided by the easy cone[24–26]. The use of ion irradiation in Ref.[23] was motivated by the fact that it can be technologically challenging to reproducibly set the easy-cone ground state in multilayer stacks containing a FeCoB/MgO interface, as it is only accessible within a narrow range of FeCoB thicknesses (see e.g. Refs.[27,28]).

However, ion irradiation has a known detrimental effect on the tunnel magnetoresistance (TMR), that has been reported for $AlO_x$-based[2,29] and MgO-based[3,30] MTJs, usually attributed to the creation of defects within the oxide barrier. While this detrimental effect may be reduced by annealing[9], it will impose an upper limit on the ion fluence that can be used to manipulate magnetic anisotropy while simultaneously keeping a functioning MTJ.

Here we explore the extent of effects produced by 400 keV $Ar^+$ irradiation on the interface-controlled parameters of a complete MTJ stack, namely, magnetic anisotropy, TMR, exchange bias, interlayer-exchange coupling and magnetization damping, by combining ferromagnetic resonance, vibrating sample magnetometry and current-in-plane tunneling techniques. The trends of the magnetic anisotropy evolution resulting from the irradiation, as well as the ion-fluence window where such a modulation of magnetic anisotropy can occur, while not losing the TMR or the magnetic configuration of the MTJ, are determined.

**Experimental details**

MTJ multilayer stacks were prepared by dc magnetron sputtering on thermally oxidized 8-inch Si wafers. The structure of the stacks is Si / SiO$_2$ / Ta(3) / CuN(30) / Ta(5) / Ru(2) / IrMn(12) / PL(2) / Ru(0.8) / RL(2.3)/ MgO(1.5) / FL(t$_{FL}$) / Ta(5) / Ru(7), where PL, RL and FL are, respectively, the pinned, reference and free layer, made of FeCo(B) alloys. The numbers in parentheses are nominal thicknesses in nanometers. The wafers were annealed at 310ºC with a 1 T applied field to set the exchange-bias direction. A total of four wafers were prepared with nominal thickness of the FL, t$_{FL}$, of 2.0 nm, 1.8 nm, 1.7 nm and 1.6 nm.

The multilayers were subsequently irradiated with $400\ \text{keV}\ \text{Ar}^+$ ions at fluences ($\Phi$) ranging from $10^{12}\ \text{cm}^{-2}$ to $5\times10^{15}\ \text{cm}^{-2}$. According to ballistic simulations using the TRIM (Transport and Range of Ions in Matter) software[31], the kinetic energy and ion mass combination guarantee that $\text{Ar}^+$ ions are implanted deep inside the Si substrate while inelastic and elastic energy-transfer processes occur within the multilayer stack. According to those simulations, elemental intermixing occurs at the interfaces, with composition changes of a few percent expected for an ion fluence of $10^{14}\ \text{cm}^{-2}$ (see SM1 of the Supplemental Material).

The extraction of magnetic parameters started from the description of the magnetic energy density, E, in $\text{mJ}\cdot\text{m}^{-2}$, of the multilayer in a macrospin approximation as

$$\begin{aligned}
E = {} & t_{FL}\left(-\mathbf{M_{FL}}\cdot\mathbf{B} - \frac{1}{2}M_{FL}B_{K1\text{eff}}^{FL}(\hat{\mathbf{m}}_{\mathbf{FL}}\cdot\hat{\mathbf{y}})^2\right) \\
& + t_{RL}\left(-\mathbf{M_{RL}}\cdot\mathbf{B} - \frac{1}{2}M_{RL}B_{K1\text{eff}}^{RL}(\hat{\mathbf{m}}_{\mathbf{RL}}\cdot\hat{\mathbf{y}})^2\right) \\
& + t_{PL}\left(-\mathbf{M_{PL}}\cdot\mathbf{B} + \frac{1}{2}\mu_0 M_{PL}^2(\hat{\mathbf{m}}_{\mathbf{PL}}\cdot\hat{\mathbf{y}})^2\right) \\
& - J_{EB}(\hat{\mathbf{m}}_{\mathbf{PL}}\cdot\hat{\mathbf{x}}) - J_{IEC}(\hat{\mathbf{m}}_{\mathbf{RL}}\cdot\hat{\mathbf{m}}_{\mathbf{PL}}) - J_{\text{Néel}}(\hat{\mathbf{m}}_{\mathbf{FL}}\cdot\hat{\mathbf{m}}_{\mathbf{RL}}),
\end{aligned} \quad (1)$$

where $\mathbf{B}$ is the external magnetic field, and $\hat{\mathbf{m}}_{\mathbf{FL}}$, $\hat{\mathbf{m}}_{\mathbf{RL}}$ and $\hat{\mathbf{m}}_{\mathbf{PL}}$ are, respectively, the unit vectors of magnetization in the FL, RL and PL with the thicknesses t$_{FL}$, t$_{RL}$ and t$_{PL}$, and with the corresponding magnetization values M$_{FL}$, M$_{RL}$ and M$_{PL}$. A scheme of the multilayer stack and the coordinate system are shown in the inset of figure 1. Both FL and RL possess an effective first-order anisotropy field, $B_{K1\text{eff}} = \left(\frac{2k_{s1}}{tM_S} - \mu_0 M_S\right)$, encompassing the competition between the interfacial perpendicular magnetic anisotropy ($k_{s1}/t$), originated at the FeCoB/MgO interfaces, and the thin-film shape anisotropy $\left(\frac{1}{2}\mu_0 M_S^2\right)$. The PL is exchange-biased to the antiferromagnetic IrMn ($J_{EB} > 0$) and coupled antiferromagnetically via a RKKY-like interlayer exchange coupling (IEC) across the Ru spacer to the RL ($J_{IEC} < 0$), constituting a typical synthetic antiferromagnet (SAF) structure. Finally, the Néel "orange-peel" magnetostatic interaction may couple the RL and the FL across the MgO, whenever there is a correlated roughness between the opposing FeCoB/MgO interfaces ($J_{\text{Néel}} > 0$).

X-band (9.87 GHz) angle-dependent ferromagnetic resonance (FMR) and vibrating sample magnetometer (VSM) measurements were carried out before and after irradiation to measure changes in $B_{K1\text{eff}}^{FL}$ and in $J_{EB}$ and $J_{IEC}$ with increasing ion fluence. As $J_{\text{Néel}} \approx 0$, the energy of the FL, $E_{FL}$, can be decoupled from that of the SAF and rewritten as:

$$E_{FL} = t_{FL}M_{FL}\left(-B\cos(\phi_B - \phi_M) - \frac{1}{2}B_{K1eff}^{FL}\sin^2\phi_M\right). \tag{2}$$

The calculated angular dependence of the resonance field (starting from equation (2) and applying the Smit-Beljers formalism[32]) was then fitted to experimental results in order to extract $B_{K1eff}^{FL}$.

Starting from equation (1), the FMR modes (precession frequency versus field) and the FMR absorption curves of the MTJ were simulated following the approaches used in Ref.[33] and Ref.[34], respectively. Angular dependences of the FMR linewidth, $\Delta B_{PP}(\phi_B)$, encompassing both the intrinsic and the inhomogeneous (due to local resonances and two-magnon scattering) broadening contributions, were simulated following the methodology of Ref.[35].

The TMR was measured via current-in-plane tunneling (CIPT) technique, with a micro 4-point prober by CAPRES A/S, following the protocol of Ref.[36]. Additionally, resistance loops as a function of field, R(H), were obtained, but for a fixed spacing of the probes. For that reason, the presented R(H) loops (see e.g. figure 4) reflect the magnetic configuration of the stack but do not reflect the magnitude of the TMR (see e.g. inset of figure 4). The CIPT technique was chosen for its flexibility, as it allowed the determination of TMR without the need to perform lateral patterning of MTJ pillars. Yet, in order to apply this technique, a metallic pathway between the prober and the magnetic free layer of the MTJ has to be ensured. Therefore, a double-MgO free layer as the one studied in Ref.[23] was not adequate for this study, and a single-MgO free layer, composed of MgO/FeCoB/Ta was used instead. This distinction in the multilayer stack design has important consequences for the effect of irradiation on the magnetic properties, that will be discussed.

## Results and discussion

### Stack properties before irradiation

Areal magnetization versus field curves (figure 1) were measured for the case of an in-plane applied external field **H**. For positive (negative) values of the field, **H** is directed antiparallel (parallel) to the exchange-bias field (i.e. $\phi_H = 180°$ for $H > 0$). Four plateaus are identified, corresponding to different magnetic configurations, depicted in figure 1 as: *(1)* saturation of the magnetization of the three layers along the field direction; *(2)* parallel configuration of the MTJ, obtained after the switching of $M_{PL}$; *(3)* antiparallel configuration of the MTJ, following the switching of $M_{FL}$; and *(4)* saturation in the direction opposite to *(1)*. Using the values of each plateau (see SM2 in the Supplemental Material) the magnetization values were estimated as $M_{RL} = 1250 \text{ kA/m}$ and $M_{PL} = 1232 \text{ kA/m}$. A $M_{FL} = 1209 \text{ kA/m}$ was estimated for the FL, presuming a 0.6 nm-thick magnetic dead layer, which is a typical value found for the MgO/FeCoB/Ta free layer[37].

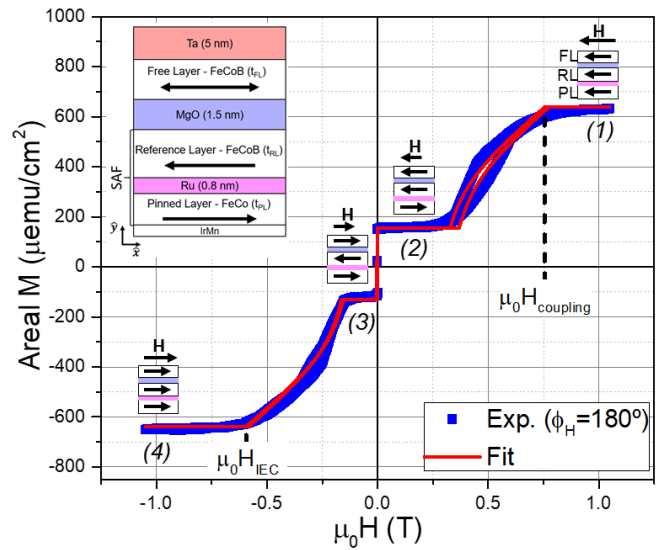

Figure 1. Magnetostatic curve of an MTJ with $t_{FL} = 2$ nm, for a magnetic field applied in plane, opposite to the exchange-bias field ($\phi_H = 180°$). Blue squares are experimental results, and the red line is a fit to the experiment following equation (1). The numbers in parentheses identify the different magnetization configurations in the MTJ, also depicted in the figure. Inset: sketch of the magnetic tunnel junction stack structure.

The interlayer exchange coupling field, $\mu_0 H_{IEC}$, is the field required to compensate the IEC when going from configuration *(3)* to configuration *(4)* in figure 1. $J_{IEC}$ is thus determined as:

$$J_{IEC} = \mu_0 H_{IEC} \frac{M_{RL} t_{RL} M_{PL} t_{PL}}{M_{RL} t_{RL} + M_{PL} t_{PL}}. \qquad (3)$$

Conversely, to achieve saturation at positive values of H, $M_{PL}$ must rotate from configuration *(2)* to an unfavorable direction, parallel to $M_{RL}$ and antiparallel to the exchange-bias field (configuration *(1)*). As a result, both the IEC and the exchange-bias coupling must be overcome by the Zeeman interaction, which occurs at the field $\mu_0 H_{coupling} = \mu_0 (H_{EB} + |H_{IEC}|)$. $J_{EB}$ is then calculated as

$$J_{EB} = \frac{1}{2}\mu_0 H_{EB} M_{PL} t_{PL}. \tag{4}$$

The mean values of the coupling constants for the four wafers were $J_{EB} = 0.459 \text{ mJ/m}^2$ and $J_{IEC} = -0.797 \text{ mJ/m}^2$. The value of $J_{IEC}$ is in accordance with that expected for the second peak of the oscillatory $J_{IEC}(t_{Ru})$ found for a 0.8 nm-thick Ru spacer[38]. Regarding a magnetostatic coupling between the FL and the RL, a field offset of $-0.2 \text{ mT}$ was registered in magnetoresistance loops (see SM4 in the Supplemental Material), which corresponds to a $J_{Néel} < 2 \times 10^{-4} \text{ mJ/m}^2$.

A simulated $M(H)$ curve, resulting from the minimization of the energy of equation (1), was fitted to the experiment. The model, based on macrospin approximation, cannot fully account for the hysteresis loop separating the configurations *(1)* and *(2)* in figure 1. The hysteresis is likely a result of a different effective $J_{EB}$ value depending on whether $H$ is decreased or increased after depinning of magnetic domains. The varying $J_{EB}$ could then be explained by a non-entirely pinned domain structure at the IrMn/PL system. Then, if the rotation of $M_{PL}$ is accompanied by expansion/contraction of domains, it will not be successfully reproduced by a macrospin model. Furthermore, small in-plane magnetic anisotropies, likely to be present in the RL and PL, will influence the shape of the loop and were not accounted for in the model. Aside from that limitation, the parameters of the fit (areal magnetization and the EB and IEC constants) agree with those obtained directly from the experimental curve, indicating the general adequacy of the model to describe the MTJ system.

The magnetic characterization of the pristine MTJs is completed by discussing the FMR results. Figure 2 contains the out-of-plane angular dependences of the FL's resonance field and the corresponding fits to the data. The FL is in-plane magnetized for all $t_{FL}$, as seen from the symmetry of the angular dependence, with a minimum (maximum) of $B_{RES}$ at $\phi_B = 0°$ ($\phi_B = 90°$). The in-plane magnetization reveals the leading role of the shape anisotropy, which results in a $B_{K1eff}^{FL} < 0$, ranging from $-0.42 \text{ T}$ for $t_{FL} = 1.6 \text{ nm}$ to $-1.07 \text{ T}$ for $t_{FL} = 2.0 \text{ nm}$ (inset of Figure 2).

Introducing a second-order anisotropy field, $B_{K2}$, did not significantly improve the quality of the fits of $B_{RES}(\phi_B)$. Thus, for the purpose of extracting $B_{K1eff}^{FL}$ we considered $B_{K2} \approx 0$. Such a second-order contribution to the anisotropy in multilayers containing FeCoB/MgO interfaces originates from an interplay between fluctuating magnetic parameters (e.g. interfacial perpendicular magnetic anisotropy) across grains and the grain-grain exchange coupling. As discussed in Ref.[27], $B_{K2}$ vanishes when the grains are either uncoupled or when the intergrain exchange is very strong. As it will be shown below, the angular dependences of the linewidth, $\Delta B_{PP}(\phi_B)$, for the FL of our MTJ reveal a small inhomogeneous broadening, mostly caused by two-magnon scattering (i.e. strongly coupled inhomogeneities). That evidence suggests the FL film to be continuous, which is consistent with a $B_{K2}$ that is much smaller than the (large) $B_{K1eff}^{FL}$.

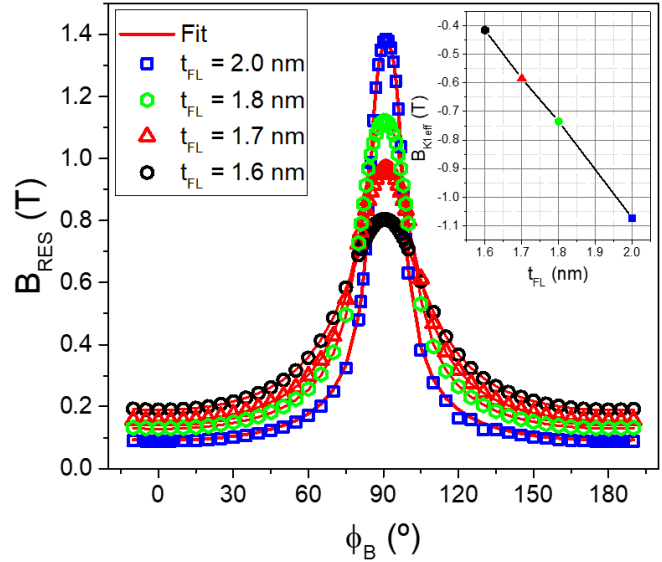

Figure 2. Angular dependence of the resonance field for $t_{FL} = 2.0$ nm (blue squares), $t_{FL} = 1.8$ nm (green hexagons), $t_{FL} = 1.7$ nm (red triangles), and $t_{FL} = 1.6$ nm (black dots). Red lines are fits to the data using the Smit-Beljers approach. Inset: FL thickness dependence of the extracted $B_{K1eff}$.

Additional resonances were observed by FMR, as shown in the top panel of figure 3(a), for **B** parallel to the exchange-bias direction ($\phi_B = 0°$). The additional FMR line is overlapped with that of the FL for $t_{FL} = 2.0$ nm, but becomes resolved for the MTJs with thinner FLs. The position of the unveiled line (around 75 mT) is independent of $t_{FL}$. The bottom panels of figure 3 contain simulated FMR spectra. A qualitative comparison between experiment and simulation allows the additional FMR line to be identified as stemming from the acoustic mode of the SAF, which is labeled as SAF-AM1. Indeed, the simulation indicates that the acoustic mode of the SAF crosses the fixed microwave frequency 4 times (see SM3 in the Supplemental Material), resulting in 4 resonances labeled SAF-AM1 through SAF-AM4. Experimentally, for $\phi_B = 0°$ it was only possible to detect SAF-AM1 and SAF-AM2. The two other modes, SAF-AM3 and SAF-AM4, were not detected. They originate absorption features with an intensity lower than SAF-AM1 and occur within the field range where the direction of $M_{RL}$ (and, through the effect of the IEC, $M_{PL}$) rotates (see M(H) in figure 1 for H < 0). Hence, the undetected modes occur in a non-saturated magnetic state, which could contribute to an inhomogeneous broadening of the absorption curves and explain why those modes were not detected. The simulated intensity of the SAF modes for $\phi_B = 180°$ (bottom panel of figure 3(b)) is nearly half of that obtained for $\phi_B = 0°$. Furthermore, in this geometry, SAF-AM modes occur in the non-saturated FMR regime (see M(H) in figure 1 for H > 0). Those reasons may explain why the SAF-AM modes are not detected at all for $\phi_B = 180°$ (top panel of figure 3(b)).

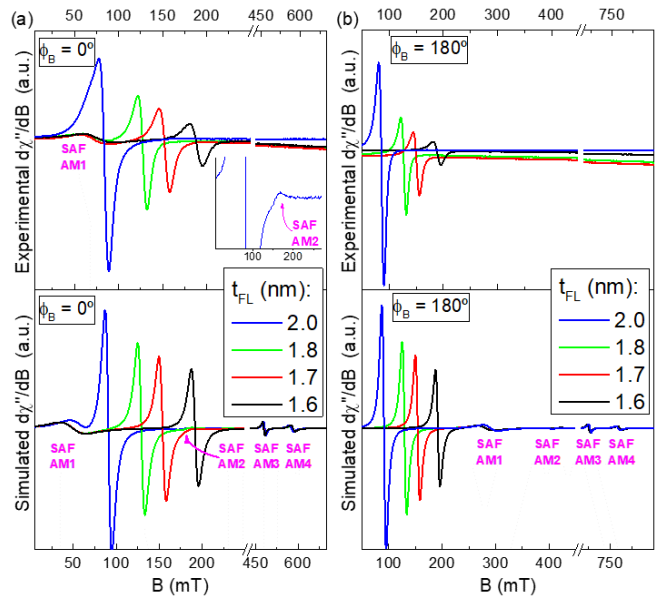

Figure 3. Experimental (top panels) and simulated (bottom panels) FMR spectra for (a) **B** parallel to the exchange-bias direction, $\phi_B = 0°$, and (b) **B** antiparallel to the exchange-bias direction, $\phi_B = 180°$, of MTJs with different free-layer thicknesses, $t_{FL}$. The intensity of the lines decreases with decreasing $t_{FL}$. The parameters of the simulation were: $B_{K1eff}^{FL}$ of figure 2, a $B_{K1eff}^{RL}= -1.6$ T (extrapolation of $B_{K1eff}^{FL}$ for a $t_{RL} = 2.3$ nm), $M_{PL} = 1232$ kA/m, $J_{EB} = 0.459$ mJ/m$^2$, $J_{IEC} = -0.797$ mJ/m$^2$, $\alpha_{FL} = 0.02$, $\alpha_{RL} = 0.04$, $\alpha_{PL} = 0.04$, g = 2.11 and f = 9.87 GHz.

The TMR increased from 127% for $t_{FL} = 1.6$ nm to 193% for $t_{FL} = 2.0$ nm, following a decrease of the resistance of the parallel state, $R_P$, while the RA product decreased from 43 $\Omega \mu m^2$ to 36 $\Omega \mu m^2$ (see SM4 in the Supplemental Material). Those tendencies suggest that the crystalline quality of the MgO barrier improves with increasing $t_{FL}$. Ta diffusion is known to hinder the coherent crystallization of FeCo/MgO required for the symmetry spin-filtering effect. One may thus expect that, the thicker the FL, the lower the concentration of the Ta diffused to the vicinity of the FeCo/MgO interface, yielding the observed improvement of TMR. For the same reason, one can expect also a slight improvement in $M_S$ near the MgO interface since Ta gets farther from it with increased thickness of the free layer.

## Stack properties after irradiation

The irradiation of the MTJ with $t_{FL} = 2.0$ nm produced a small decrease in the total magnetization of the stack and of the coupling fields up to a fluence of $3 \times 10^{13}$ cm$^{-2}$, as seen in the VSM results of figure 4(a). In contrast, the irradiation with $\Phi > 10^{14}$ cm$^{-2}$ resulted in a significant change in the shape of the magnetization curve, characterized by a decrease in the total $M_S$ accompanied by a decrease of both coupling fields down to $\mu_0 H_{IEC} \approx 0.1$ T and $\mu_0 H_{EB} \approx 0.05$ T. Ultimately, for a $\Phi = 10^{15}$ cm$^{-2}$, a single switching event for the whole MTJ is observed at $H \approx 0$. The decrease in the coupling energies with increasing irradiation fluence is presented in figure 4 (b).

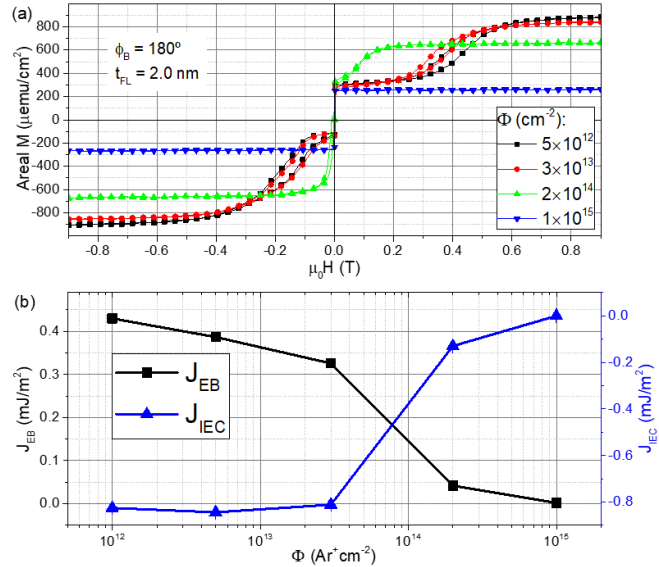

Figure 4. (a) Magnetostatic curve of Ar$^+$ irradiated MTJs with a $t_{FL} = 2$ nm. (b) Dependence of the coupling energies, $J_{EB}$ and $J_{IEC}$, on the ion irradiation fluence.

With increasing irradiation fluence, the resonance field of the FL's FMR spectra increased (decreased) for an in-plane (out-of-plane) applied magnetic field as shown in figure 5, where, for each $\Phi$, the field axis was normalized by the resonance field of the FL in the corresponding pristine sample, $B_0^{non-irrad}$. That was done in order to account for the (small) variability of anisotropy between different pieces of the wafer. The progression of the FL's peak position is consistent with a decrease in the magnitude of $B_{K1eff}^{FL}$, i.e. the anisotropy keeping the magnetization in plane is reduced by increasing fluence. Indeed, as figure 6(a) shows, the anisotropy field after irradiation, relative to the anisotropy of the pristine sample, $B_{K1eff}^{\Phi}/B_{K1eff}^{0}$, follows an exponential decay (please note the logarithmic scale of the x-axis). The characteristic fluence of the decay is of about $10^{14}$ cm$^{-2}$ for the MTJ with $t_{FL} = 2.0$ nm, and it decreases for the thinner layers. In other words, the irradiation-induced modulation of the magnetic anisotropy is more pronounced for the thinner layers. The characteristic $\Phi$ is also evident in the spectra of figure 5: the shift in resonance field becomes more pronounced and the amplitude of the FMR line gets much smaller above $10^{14}$ cm$^{-2}$.

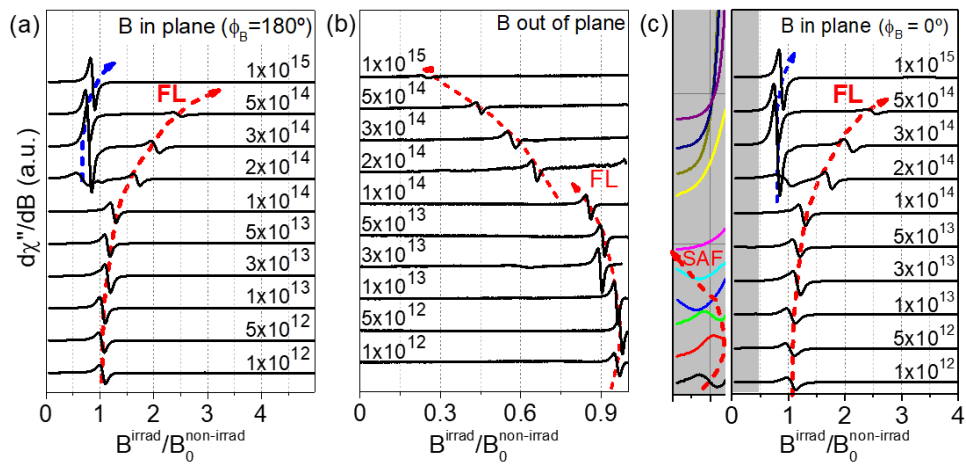

Figure 5. FMR spectra of irradiated MTJ with a $t_{FL} = 2.0$ nm for (a) **B** in plane and antiparallel to exchange-bias direction, (b) **B** perpendicular to the layers, and (c) **B** in plane and parallel to the exchange-bias direction. The ion fluence increases from the bottom to the top and is indicated atop each spectrum. The field axis has been normalized by the resonance field of the FL in the pristine MTJs, $B_0^{non-irrad}$.

It is interesting to note that the evolution of the anisotropy with increasing fluence for a FL composed of MgO/FeCoB/Ta occurs with an opposite trend to that observed for a double-MgO free layer, MgO/FeCoB/MgO, as seen in Ref.[23]. The difference is explained by the distinct surroundings of the FeCoB layers. In the double-MgO layer, ion-induced intermixing of the FeCoB/MgO interface produces a more rapid decrease of $k_{s1}$ than of $M_S$, resulting in a $B_{K1eff}^{FL}$ evolving into the negative values. Consequently, in double-MgO free layers, with increasing ion fluence, it is possible to reorient the magnetization from easy axis to easy cone and then to easy plane[23]. In contrast, in a Ta-capped free layer, the ion irradiation promotes a stronger decrease of $M_S$ due to intermixing at the top FL/Ta interface than of $k_{s1}$ at the MgO/FeCoB and, thus, $B_{K1eff}^{FL}$ evolves in the direction of positive values, as schematized in figure 6 (b).

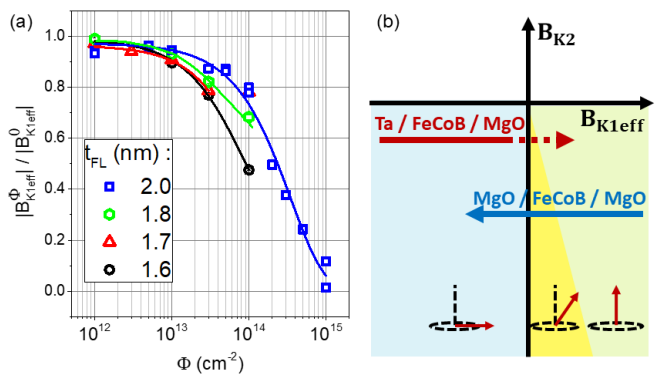

Figure 6. (a) Relative change in magnitude of the FL's effective first-order anisotropy field, $B_{K1eff}^{\Phi}$, as a function of the ion irradiation fluence, for $t_{FL} = 2.0$ nm (blue squares), $t_{FL} = 1.8$ nm (green hexagons), $t_{FL} = 1.7$ nm (red triangles) and $t_{FL} = 1.6$ nm (black circles). The reference mean values of $B_{K1eff}^{0}$ for the pristine MTJs are indicated in figure 2. (b) Sketch of the evolution of $B_{K1eff}^{\Phi}$ with increasing ion fluence for the case of a Ta-capped FL (red arrow) and for a double-MgO FL (blue arrow).

By extrapolating the trend observed for the anisotropy of the Ta-capped free layer, the easy-cone anisotropy is expected to be reached by irradiating an initially in-plane magnetized film (see figure 6(b)). Yet, in the MgO/FeCoB/Ta multilayers investigated here, the easy cone was not attained (at room temperature) with increasing ion fluence. There are at least two reasons for that: First, $B_{K2}$ in these Ta-capped films is smaller in comparison with the double-MgO free layers of Ref. [23], as discussed above. Starting with a smaller $|B_{K2}|$ restricts the range of $B_{K1eff}$ values for which the easy cone can be obtained. Secondly, in Ta-capped films, the physical mechanism behind the modulation of anisotropy is also at the basis of the loss of ferromagnetic order: $M_S$ is decreased following the intermixing at the FL/Ta interface. Nonetheless, the temperature can be explored to reach the easy cone, taking advantage of the fact that, upon cooling, $B_{K1eff}$ is expected to increase into the positive range and $|B_{K2}|$ to scale with $(B_{K1eff})^2$, as pointed out in Ref.[28]. Thus, starting with the multilayer irradiated with $10^{15}$ cm$^{-2}$, which exhibits a $B_{K1eff} \approx 0$ (see figure 6(a)), i.e. near the crossover from in-plane to easy-axis anisotropy, and decreasing the temperature to 150 K, the easy cone could be reached, as presented in figure 7. For future developments of the ion-irradiation-induced easy-cone anisotropy in single-MgO free layers, the exploration of different elements to be used as capping is of potential interest. For instance, replacing the capping of Ta by W, one can expect a smaller decrease of $M_S$ upon ion irradiation, due to a smaller degree of intermixing at the FL/W than at the FL/Ta interface. Those differences may be explained by a different miscibility of W and of Ta in FeCo (see, e.g. Ref.[39]). Indeed, in the double-MgO free layers having those elements as spacers (see Ref. [23]), one and the same ion fluence produced a change in the easy-cone angle in the FL with a W spacer, whereas it leads to a complete reorientation of the magnetization direction from easy axis to easy cone and then to easy plane in the FL with a Ta spacer.

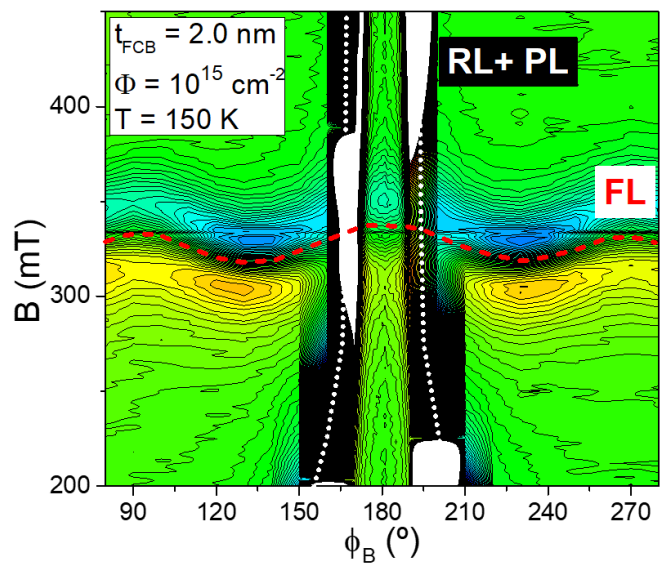

Figure 7. Contour plot of the angular dependent FMR spectra obtained at 150 K for a MTJ with a MgO / FeCoB (2 nm) / Ta free layer irradiated with $\Phi = 10^{15}$ cm$^{-2}$. The color scale represents the peak amplitude of the FMR spectra, in arbitrary units, with peak maxima in orange and minima in black. The red dashed line marks the approximate resonance field of the FL, while the white dashed line indicates the peak position of the ferromagnetically coupled reference layer and pinned layer (RL+PL) due to irradiation-induced intermixing of the SAF (see discussion in the text and figures below).

The effects of the ion irradiation on the layers composing the SAF were also evaluated in a qualitative manner. In figure 5(c) it can be seen that, below $10^{14}$ cm$^{-2}$, the SAF-AM1 mode shifts to lower fields with increasing fluence. That shift is attributed to the initially small decrease in $J_{EB}$ (see figure 4(b)), according to simulations (SM5 in the Supplemental Material).

For fluences higher than $10^{14}$ cm$^{-2}$, $J_{EB}$ and $J_{IEC}$ get vanishingly small, and consequently the magnetic configuration typical of a SAF is no longer maintained: the PL becomes decoupled from the IrMn antiferromagnet, and the antiferromagnetic-like IEC between RL and PL is lost. In fact, above $2 \times 10^{14}$ cm$^{-2}$ it is reasonable to presume that the RL and PL become ferromagnetically coupled and behave as a single, thicker ferromagnetic layer. The loss of the exchange bias and a ferromagnetic coupling between RL and PL explain the appearance of only one additional FMR line in figure 5(a, c).

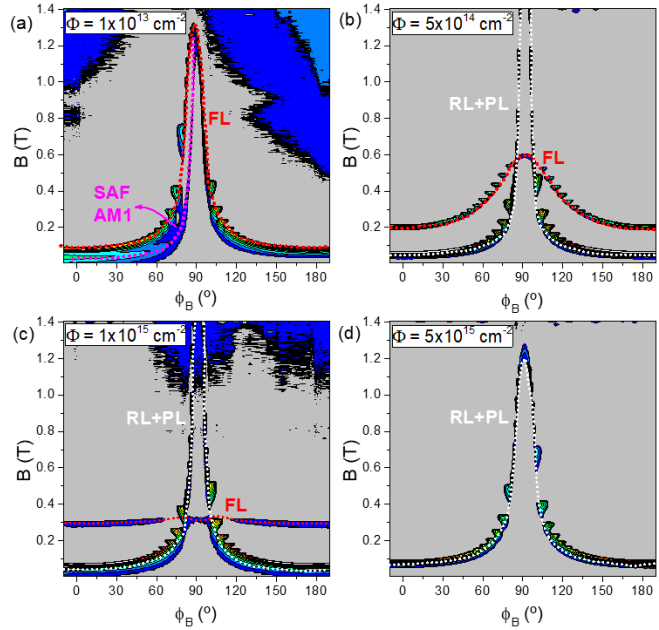

Figure 8. Angular dependences of the resonance fields measured for MTJs irradiated with: (a) $\Phi = 10^{13}$ cm$^{-2}$; (b) $\Phi = 5 \times 10^{14}$ cm$^{-2}$; (c) $\Phi = 10^{15}$ cm$^{-2}$; and (d) $\Phi = 5 \times 10^{15}$ cm$^{-2}$. The color code represents the amplitude of the positive peak of the FMR intensity, d$\chi''$/dB, in arbitrary units. Dotted lines indicate the evolution of $B_{RES}(\phi_B)$ for the cases of the FL (red), SAF-AM1 (pink) and RL+PL (white).

The angular dependences of the resonance field presented in figure 8 summarize the referred events that take place with increasing fluence. In the low-fluence range, as at $\Phi = 10^{13}$ cm$^{-2}$ (figure 8(a)), the anisotropy field of the FL is nearly the same as the one in the pristine MTJ, and a resonance identified as SAF-AM1 is observed for $\phi_B < 90°$. In the high-fluence range, namely at $\Phi = 5 \times 10^{14}$ cm$^{-2}$ (figure 8(b)), the $B_{K1eff}^{FL}$ drops below 50% of the corresponding value for the pristine sample. Additionally, a resonance having an angular dependence that reflects a $B_{K1eff} < 0$ becomes apparent. That anisotropy field is stronger in magnitude than the one of the FL, i.e. $|B_{K1eff}| > |B_{K1eff}^{FL}|$, as indicated by a peak position that surpasses the experimental field range around $\phi_B = 90°$. The high effective magnetic moment associated with that strong anisotropy is likely to be a result of the ferromagnetic coupling between the PL and the RL. That coupling probably occurs due to the ion-induced intermixing of the PL/Ru/RL interfaces, which reduces the effective thickness of the Ru spacer. However, it is not clear whether the effective $t_{Ru}$ decreases towards a value where $J_{IEC}$ becomes positive or if the intermixing occurs to an extent where direct interfacial exchange coupling between RL and PL is promoted. After irradiation with $\Phi = 10^{15}$ cm$^{-2}$ (figure 8(c)), the FL gets practically magnetically isotropic, with the resonance occurring at the field value corresponding to $g \approx 2$, which is indicative of the onset of a ferromagnetic-paramagnetic transition in that layer. Ultimately, at $5 \times 10^{15}$ cm$^{-2}$ (figure 8(d)), the FL absorption is no longer observed, and $B_{K1eff}$ of the FM-coupled RL and PL decreases.

The results also show in a qualitative manner that, apart from the coupling energies, the magnetic properties of the SAF are more robust to the irradiation than those of the FL: the decrease in FMR line intensity and in $B_{K1eff}$ happens at higher fluences for the layers composing the SAF than for the FL. Considering the high energy of the incident ions and the small spatial separation between FL and SAF (they are separated only by a 1.5 nm MgO barrier), the energy density deposited in the SAF is practically identical to that deposited in the FL. The different extent of irradiation-produced effects in the two layers is thus rather explained by their distinct layer surroundings: the (thinner) FL is in contact with a thick Ta capping layer, while the (thicker) SAF is enclosed between the MgO barrier and the IrMn layer, only interrupted by a thin Ru spacer. A lower average threshold displacement energy of Ta than of Ru (Ref.[40]) may favor a higher degree of intermixing at the FL/Ta than at the PL/Ru/RL interfaces. Furthermore, the higher effective thickness of the RL+PL system should also contribute to the preservation of the FMR peak intensity up to higher fluences (see figure 8(d)). The intermixing of the layers upon ion irradiation is qualitatively corroborated by XRR results (SM6 of the Supplemental Material).

The characteristic fluence of $10^{14}$ cm$^{-2}$, seen in the VSM and FMR results, is reflected also in the electrical properties of the irradiated MTJs. Below $10^{14}$ cm$^{-2}$, the TMR drops down to a value of 74% at $\Phi = 3 \times 10^{13}$ cm$^{-2}$, following the decrease in the resistance of the antiparallel state, R$_{AP}$ (figure 9). The intermixing at the MgO/FeCoB interfaces cannot explain that initial loss of TMR, since the resistance of the parallel state, R$_P$, and $B_{K1eff}^{FL}$ remain practically unchanged up to $\Phi = 3 \times 10^{13}$ cm$^{-2}$. A possibility is instead the creation of defects within the MgO barrier. Those defects may constitute an additional electron tunneling channel, characterized by a resistance $R_D$. If the magnitude of $R_D$ is comparable to the resistance of the pristine MTJ in the AP state, $R_{AP}^0$, then R$_{AP}$ will decrease after irradiation as a result of the parallel connection of $R_{AP}^0$ and $R_D$. On the other hand, the R$_P$ value would remain practically unchanged as long as $R_D \gg R_P^0$. Above $\Phi = 10^{14}$ cm$^{-2}$, the magnetization of the RL is no longer pinned due to the loss of the magnetic coupling, resulting in an undefined antiparallel state, as seen by the emergence of two loops in the R(H) curve. Additionally, R$_P$ is significantly decreased, suggesting an irradiation-induced deterioration of the crystallinity of the MgO barrier and of the MgO/FeCoB interfaces. At $10^{15}$ cm$^{-2}$, the R(H) dependence becomes flat and the TMR goes to zero.

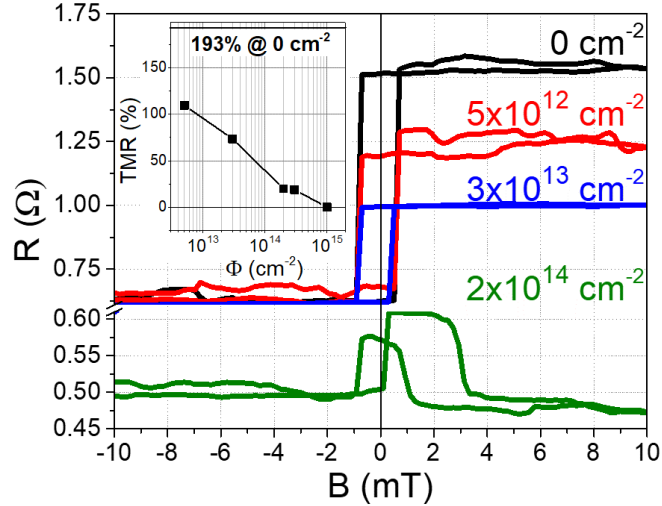

Figure 9. Resistance versus field loops for an MTJ stack before irradiation (black) and after irradiation with $\Phi = 5 \times 10^{12}$ cm$^{-2}$ (red), $\Phi = 3 \times 10^{13}$ cm$^{-2}$ (blue), and $\Phi = 2 \times 10^{14}$ cm$^{-2}$ (green). Inset: dependence of the TMR on the irradiation fluence. The solid black line indicates the TMR of the pristine MTJ stack, at a value of 193%.

At last, the impact of the irradiation on the magnetization dynamics was analyzed by fitting the angular dependence of the FMR linewidth, ($\Delta B_{PP}(\phi_B)$ - see figure 10), with a model comprising the intrinsic broadening and the inhomogeneous broadening caused by fluctuating anisotropy and two-magnon scattering. For the cases of pristine MTJ and low-fluence irradiation (figure 10(a)) the model adequately describes the experimental results, confirming the intrinsic damping as the dominant contribution to the line broadening. For fluences higher than $10^{14}$ cm$^{-2}$ (figure 10(b)), the shape of the $\Delta B_{PP}(\phi_B)$ evidences the increase in inhomogeneous broadening, as one would expect for a significantly damaged magnetic layer. Furthermore, the model no longer produces a satisfactory fit to the data, reason why the damping parameters are not extracted for fluences above $10^{14}$ cm$^{-2}$. The inadequacy of the model may be linked to an increase in $J_{Néel}$, with the coupling between FL and RL leading to a distorted angular dependence of the linewidth in a way that is not predicted by the model.

Interestingly, different damping constants are found whether the projection of the field on the MTJ plane is parallel ($\phi_B = 0°$) or antiparallel ($\phi_B = 180°$) to the exchange-bias direction. In other words, there is a linewidth asymmetry about $\phi_B = 90°$. The fit to $\Delta B_{PP}(\phi_B)$ thus yields two different damping values: $\alpha$ for $\phi_B > 90°$ and $\alpha'$ for $\phi_B < 90°$, with $\alpha' > \alpha$. The linewidth asymmetry is quantified through the ratio $\Delta B_{PP}(\phi_B = 0°)/\Delta B_{PP}(\phi_B = 180°)$ and plotted against the ion fluence in figure 11(a). The asymmetry exists already in the non-irradiated stack and then increases with the fluence, peaking at $10^{13}$ cm$^{-2}$, before disappearing above $10^{14}$ cm$^{-2}$. Figure 11(b) shows the changes in the damping constants, relative to the value obtained for each sample before irradiation, i.e. $\alpha(\Phi) - \alpha(0)$. No clear dependence of $\alpha$ on the irradiation fluence is found, i.e. $\alpha$ is not significantly impacted by the irradiation at low fluences ($\Phi_B < 10^{14}$ cm$^{-2}$). On the other hand, the change in $\alpha'$ peaks at $\Phi_B = 10^{13}$, in correlation with the linewidth asymmetry.

Bruno Teixeira          09/04/2020

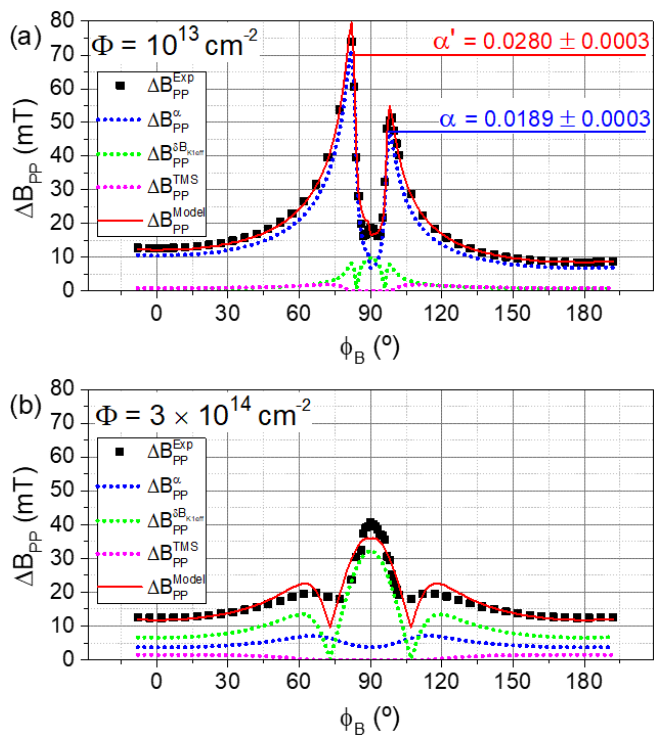

Figure 10. Angular dependences of the FL's peak-to-peak linewidth for MTJ stacks irradiated with (a) $\Phi = 10^{13}$ cm$^{-2}$ and (b) $\Phi = 3 \times 10^{14}$ cm$^{-2}$. Black squares are experimental results and the red solid lines are fits to the data using a model that encompasses the intrinsic broadening (blue dotted line), spatial fluctuations of $B_{K1eff}^{FL}$ (green dotted line) and two-magnon scattering (pink dotted line) contributions. In (a) and (b), a higher damping constant is found for $\phi_B < 90°$ ($\alpha'$) than for $\phi_B > 90°$ ($\alpha$).

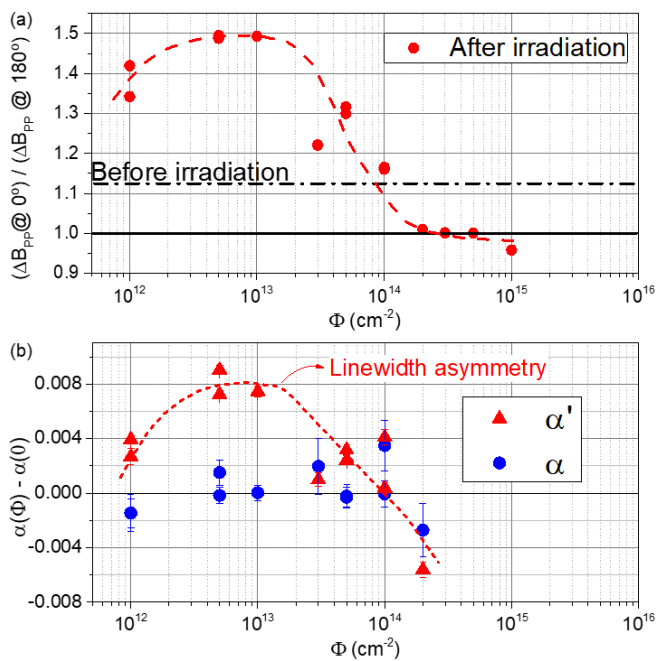

Figure 11. (a) Angular asymmetry of the linewidth, defined as the ratio between $\Delta B_{pp}(\phi_B = 0°)$ and $\Delta B_{pp}(\phi_B = 180°)$, versus the irradiation fluence. The horizontal dashed black line indicates the asymmetry for the pristine sample. The dashed red curve is a guide-to-the-eye of the asymmetry evolution with the fluence. (b) Changes in the magnetic damping, relative to the pristine sample, defined as $\alpha(\Phi) - \alpha(0)$, for the damping obtained for $\phi_B < 90°$ ($\alpha'$ - red triangles), and for the damping obtained for $\phi_B > 90°$ ($\alpha$ - blue dots).

One potential explanation for the linewidth asymmetry would be the hybridization of the FMR modes, whereby, through the Néel coupling, the resonance in the FL would be perturbed by the resonance in the SAF. The degree of hybridization and the corresponding line broadening would then depend on the different magnetic configurations of the MTJ (either P or AP) due to the different separation between the resonances in the FL and in the SAF. In our case, $J_{Néel}$ is rather small ($J_{Néel} < 2 \times 10^{-4}$ mJ/m$^2$) and for fluences below $10^{14}$ cm$^{-2}$ the effective Néel coupling field acting on the FL does not change, as revealed by the constant shift of the FL's hysteresis loop seen in the VSM and CIPT results (e.g. figure 9). A simulation of the absorption curves (SM7 of the Supplemental Material) confirmed the $J_{Néel}$ value to be insufficient to account for the linewidth asymmetry.

Notably, the higher damping, $\alpha'$, is measured for the AP state of the MTJ, whose resistance decreases with increasing fluence (see figure 9). Furthermore, for fluences above $10^{13}$ cm$^{-2}$, the linewidth asymmetry, that is correlated with $\alpha'$, starts to decrease until it completely disappears around $10^{14}$ cm$^{-2}$. That decrease in the linewidth asymmetry can be ascribed to the fact that the AP state of the MTJ is no longer maintained at the magnetic fields for which the resonance occurs (see e.g. loss of AP configuration above 4 mT for a $\phi = 2 \times 10^{14}$ cm$^{-2}$ irradiated stack in figure 9). The correlation between linewidth asymmetry, magnetic configuration of the MTJ and resistance suggests that a spin pumping effect is at play, contributing to the value of $\alpha'$. In that case, $\alpha' = \alpha + \alpha_{SP}$, yielding a spin-pumping contribution of $\alpha_{SP} = 0.0091$, for the MTJ irradiated with $10^{13}$ cm$^{-2}$. That would correspond to a spin mixing conductance, $G^{\uparrow\downarrow} = \frac{4\pi t_{FL} M_{FL}}{\mu_B g} \alpha_{SP} \approx 10^{15}$ cm$^{-2}$. The order of magnitude of the estimated $G^{\uparrow\downarrow}$ is typical of metallic interfaces[34] and thus unreasonably high for the case of an MgO barrier, which is expected to partially suppress spin pumping[41]. We tentatively attribute the estimated $G^{\uparrow\downarrow}$ to an anisotropic spin-pumping at the FL/Ta interface. That anisotropic source of damping from the FL/Ta interface would depend on the spin currents being emitted across the MgO barrier, which in turn depends on the magnetic configuration of the MTJ and on its electrical resistance, both of which are impacted by the irradiation. The exact mechanism at the basis of such an anisotropic spin-pumping is, however, not understood and will required further investigation.

## Conclusions

Combining FMR, VSM and CIPT techniques, the effects of 400 keV Ar⁺ ion irradiation on the magnetic and electrical properties of MTJ stacks were tracked. A correlation was found between the fluence-dependent changes in magnetic anisotropy, coupling energies, TMR and damping, which allowed to distinguish between two irradiation regimes.

In the low-fluence regime, $\Phi < 10^{14}$ cm$^{-2}$, there is a weak modulation of the free-layer anisotropy due to a decrease in $M_S$, and there is a continuous decrease of the TMR down to about 70% at $\Phi = 3 \times 10^{13}$ cm$^{-2}$. The drop in TMR is due to a decrease in $R_{AP}$, likely caused by the creation of defects inside the MgO barrier, which act as spin-independent tunneling channels in parallel with the spin-dependent one. The interlayer exchange coupling and the exchange bias decrease slightly, but the SAF magnetic structure is preserved. No significant changes in the magnetization damping are observed.

In the high-fluence regime, $\Phi > 10^{14}$ cm$^{-2}$, the modulation of the free-layer anisotropy is strong. It is caused by a strong decrease in $M_S$, likely due to a high degree of elemental intermixing at the FL/Ta interface. Ultimately, irradiation at fluences around $\Phi = 10^{15}$ cm$^{-2}$ renders the FL paramagnetic. The intermixing of the PL/Ru/RL and of the IrMn/PL interfaces results in a loss of $J_{EB}$ and of $J_{IEC}$. As a consequence, the magnetization of the RL is no longer pinned and the antiparallel state of the MTJ cannot be maintained. The TMR vanishes due to the damaged MgO barrier and MgO/FeCoB interfaces.

The results show, through the various parameters relevant to MTJ applications, that there is a window of operation in what concerns the use of ion irradiation for the tailoring of magnetic anisotropy. For the ion energy and mass used in this study and for the typical thicknesses found in MTJ stacks, that window is limited by a characteristic fluence of the order of $10^{14}$ cm$^{-2}$. Below that fluence, small changes in anisotropy can be induced at the cost of negatively impacting other interface-controlled parameters of the MTJ, namely TMR, but still keeping a functional MTJ. On the contrary, above that fluence the MTJ is rendered inoperative.

It was further demonstrated that the layers surrounding the magnetic free layer play a decisive role in determining the trend of the ion-irradiation-induced magnetic anisotropy modulation. If a FeCoB layer is sandwiched between two MgO layers (double-MgO), intermixing at the interfaces promotes a decrease in interfacial PMA (Ref.[23]) that leads to a reorientation of magnetization in the direction from perpendicular easy axis to easy cone and then to easy plane with increasing ion fluence. On the contrary, if the FeCoB is capped by a metal such as Ta, intermixing at the FeCoB/Ta interface reduces the effective magnetization and can promote a reorientation from easy plane to easy cone or even easy axis. However, there seems to be a fine balance between decreasing $M_S$ by a sizeable amount to induce those reorientations but not too much as to lose ferromagnetic order. The use of alternative capping materials, such as W, which are less prone to diffuse through FeCo, is suggested for future research regarding anisotropy modulation via ion irradiation.

## Acknowledgements

Work developed within the scope of the projects i3N, UIDB/50025/2020 & UIDP/50025/2020, financed by national funds through the FCT/MEC. B.M.S.T. acknowledges financial support by FCT through the bursary PD/BD/113944/2015 and BI-52 (33403/2019). B.M.S.T. and N.A.S. were supported by the European Project H2020 – MSCA – RISE – 2017 – 778308 – SPINMULTIFILM.

Bruno Teixeira                                                                                          09/04/2020

**Supplemental Material: Ion beam modification of magnetic tunnel junctions**

B. M. S. Teixeira[1,*], A. A. Timopheev[2], N. Caçoilo[1,3], L. Cuchet[2], J. Mondaud[2], J. R. Childress[2], S. Magalhães[4], E. Alves[4], N. A. Sobolev[1]

[1] i3N, Departamento de Física, Universidade de Aveiro, 3810-193, Aveiro, Portugal

[2] Crocus Technology, 3 avenue Doyen Louis Weil, BP1505 - 38025 GRENOBLE Cedex1, France

[3] Univ. Grenoble Alpes, CEA, CNRS, Grenoble INP, SPINTEC, F-38000 Grenoble, France

[4] IPFN, Instituto Superior Técnico, Universidade de Lisboa, 2695-066 Bobadela LRS, Portugal

* Corresponding author: bmsteixeira@gmail.com

## SM1. Simulation of ion irradiation of MTJ

The simulations were carried using the TRIM (Transport and Range of Ions in Matter) software[31]. In the simulations the multilayer stack is assumed to be at 0 K, which means there is no thermal diffusion of atoms or the self-annealing typical of metals. Also, there is no build-up of damage, meaning that each ion sees a pristine target. This leads to an underestimation of the produced damage. Moreover, the crystalline structure is not considered in the simulations. Nevertheless, the assumptions are not expected to significantly impact the (qualitative) goal of the simulations which was to guarantee that the Ar$^+$ ions are implanted inside the Si substrate (figure S1), whereas elastic and inelastic interactions occur within the multilayers, with expected intermixing of elements across the interfaces (figure S2). As seen in figure S1, at 400 keV the peak of Ar$^+$ implantation is located well inside the Si substrate. For a fluence of $10^{14}$ cm$^{-2}$, the volume concentration of Ar$^+$ inside the multilayer stack is smaller than $6 \times 10^{17}$ Ar$^+ \cdot$ cm$^{-3}$, i.e. 5 orders of magnitude smaller than the atomic density of the elements comprising the layers.

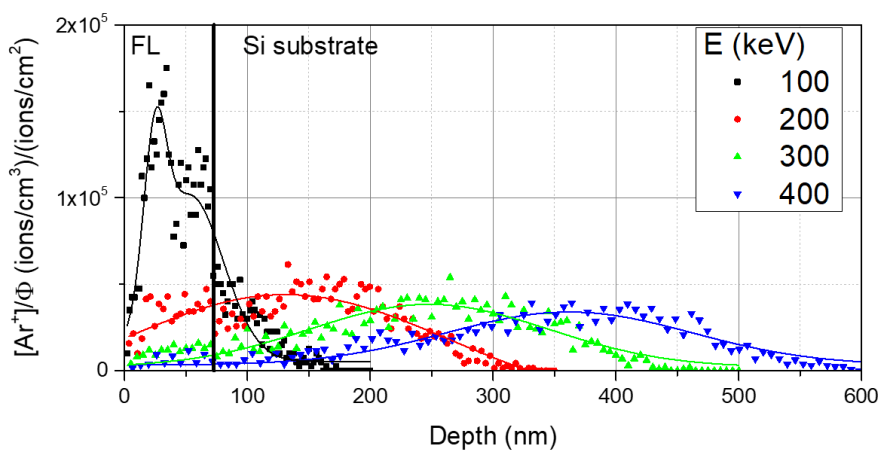

Figure S1. Simulated concentration profile of Ar$^+$ ions, normalized by the ion fluence, $[\mathrm{Ar}^+]/\Phi$, as a function of depth in the MTJ stack deposited on Si substrate for different ion energies: 100 keV (black squares); 200 keV (red dots); 300 keV (green up triangles) and 400 keV (blue down triangles). Solid lines are guides-to-the-eye of the ion range distribution. The vertical line indicates the depth corresponding to the substrate surface. The concentration of implanted Ar$^+$ at each depth can be obtained by multiplying the respective y-axis value by the ion fluence $\Phi$.

The changes in the element concentration profiles, in units of percentage per fluence (%$\Phi^{-1}$), were also calculated, assuming a regime of linear mixing. In that regime, the probability of the same atom to be recoiled twice is negligible. Within that assumption, the change in concentration of the elements of the stack after the irradiation is simply the difference between the number of atoms recoiled into vacancies (r) and the created vacancies (v). After irradiation with a fluence Φ, the percentual composition variation, in a given layer of atomic density ρ, is calculated as:

$$\text{Concentration changes} = 100\,\Phi\,\frac{(r-v)}{\rho}. \quad (1)$$

The expected composition variations are shown in figure S2, with positive (negative) values representing the accumulation (depletion) of elements relative to the pristine sample. According to the simulations, concentration changes of the order of a few per cent are expected for a $10^{14}$ cm$^{-2}$ fluence.

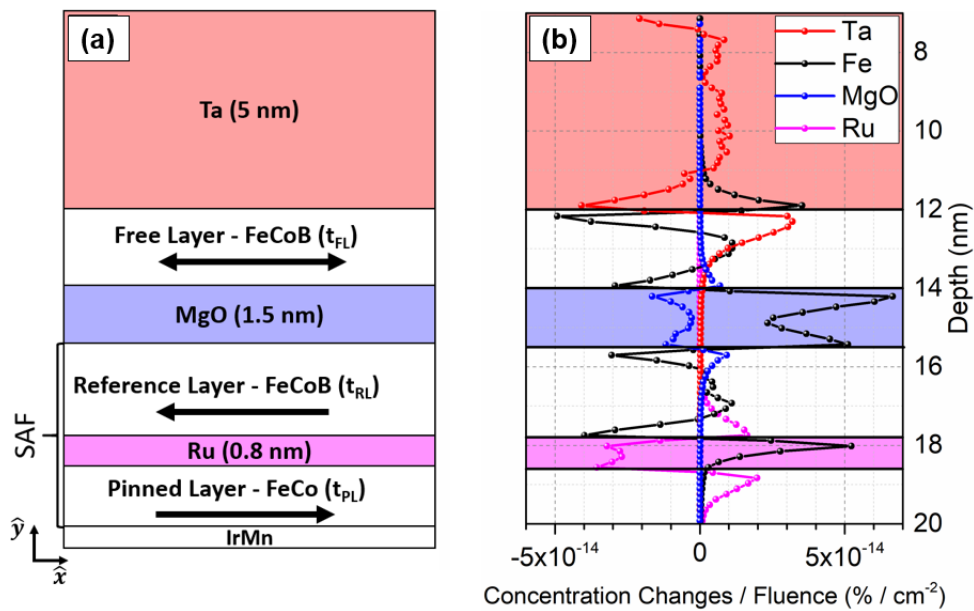

Figure S2. (a) Magnetic tunnel junction stack composed of free layer, FL, reference layer, RL, and pinned layer, PL; (b) TRIM-simulated changes to the elemental concentration (in units of percentage per ion fluence – see equation 1) along the depth of the multilayer stack presented in (a).

## SM2. Estimate of magnetic parameters from vibrating sample magnetometry results

The mean values of each plateau (respectively Mt1, Mt2 and Mt3) of figure S3 (a) were used to estimate the areal magnetization of each layer, by solving the following system of equations:

$$\begin{pmatrix} 1 & 1 & 1 \\ 1 & 1 & -1 \\ -1 & 1 & -1 \end{pmatrix} \begin{pmatrix} M_{FL}t_{FL} \\ M_{RL}t_{RL} \\ M_{PL}t_{PL} \end{pmatrix} = \begin{pmatrix} Mt1 \\ Mt2 \\ Mt3 \end{pmatrix}. \qquad (2)$$

The solutions to equation 2 are shown in figure S3 (b), as a function of the free-layer thickness. As expected, $M_{RL}t_{RL}$ and $M_{PL}t_{PL}$ show no dependence on $t_{FL}$. Assuming the nominal values of $t_{RL}$ and $t_{PL}$, i.e. considering there is no magnetic dead layer in RL and PL, the magnetization saturation values are estimated to be 1250 kA/m and 1232 kA/m, respectively. Also as expected, $M_{FL}t_{FL}$ evolves linearly with $t_{FL}$. As the stack measured in VSM was cut from a wafer, the sample area could not be determined with a sufficient accuracy to precisely determine the magnetic dead layer thickness of the FL (i.e. $t_{FL}$ yielding $M \times t = 0$ in figure S3 (b)). Indeed, assuming a (modest) uncertainty of 10% in the sample area, the magnetic dead layer thus calculated would fall within a broad interval: $0.4 < t_{dead} < 1.4$ nm. Presuming a $t_{dead} = 0.6$ nm, which is a typical value found in literature for the MgO/FeCoB/Ta free layer [37], a $M_{FL} = 1209$ kA/m is estimated.

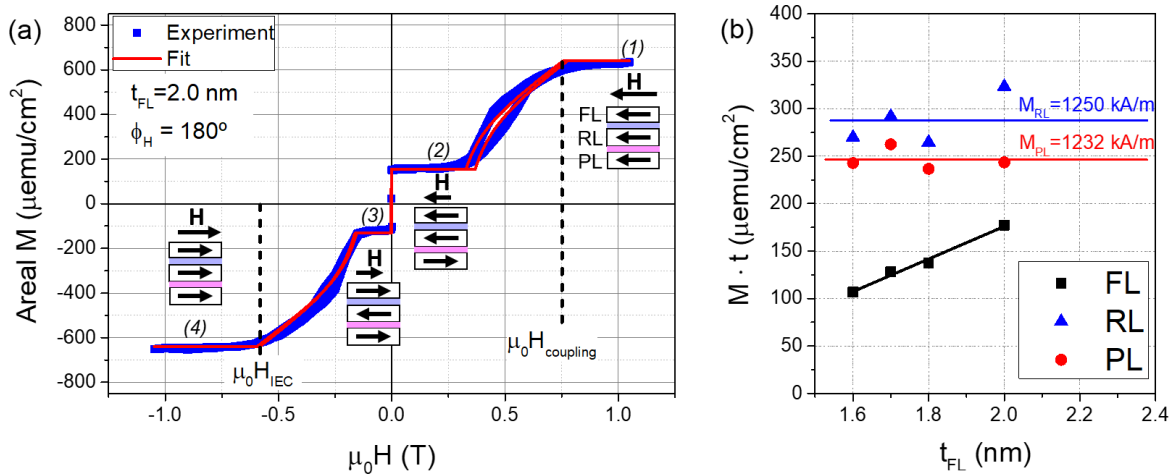

Figure S3. (a) Magnetostatic curve of an MTJ with $t_{FL} = 2\ nm$, for a magnetic field applied in plane, opposite to the exchange-bias field ($\phi_H = 180°$). Blue squares are experimental results and the red line is a fit to the experiment. Numbers in parenthesis identify the different magnetization configurations in the MTJ, also depicted in the figure. (b) Dependence of the areal magnetization of the free layer, FL, reference layer, RL, and pinned layer, PL, on the thickness of the free layer, $t_{FL}$. Horizontal lines indicate the mean values of $M_{RL}t_{RL}$ and $M_{PL}t_{PL}$, while the black line is a fit to the $M_{FL}t_{FL}(t_{FL})$ dependence (black squares).

## SM3. FMR modes for a magnetic tunnel junction

Figure S4 shows simulation results for the case of an external field applied along the exchange-bias direction (along the x axis) of the MTJ. The parameters of the simulations are included in the figure caption. The first panel contains the magnetostatic curve, obtained from the projection of $\mathbf{M_i} = M_{S_i}\mathbf{m_i}$ onto $\mathbf{B}$. Two configurations are found at the plateaus of the M(B) curve: the saturation of all layers along B at high fields *(1)*, and the antiparallel state of the MTJ *(2)*. The rotation of $M_{RL}$ starts at B ≈ 0.5 T, the point at which minimization of IEC energy, rather than minimization of the Zeeman energy, brings about a greater reduction in the magnetic energy density. As $J_{IEC} > J_{EB}$, it is energetically favourable for $M_{PL}$ to be "dragged" along (in the opposite direction) by the rotating $M_{RL}$. Figure S4 (c) shows how $M_{PL}$ rotates in the sample plane out of the x-axis by up to 45⁰ (at a field of 0.25 T) before rotating back to its original position.

The FMR modes are shown in the second panel of Figure S4 (a). The FL, under the effect of a vanishingly small $J_{Néel}$, presents the f(B) dependence typical of the Kittel equation. In contrast, the dynamical responses of $M_{RL}$ and $M_{PL}$ are strongly coupled by the IEC, to the extent that it is only meaningful to speak about the dynamics of the whole SAF. The FMR response of the SAF is then described by two normal modes: the acoustic mode, AM, of the lower-frequency and in-phase precession of the magnetic moments in each layer; and the optical-mode, OM, of the higher-frequency and out-of-phase precession. The experimental frequency $f_0$, included in the figure as a dashed line, crosses the SAF-AM four times and the FL mode once, at specific field values – the resonance fields, $B_{RES}$. The field derivate of $Im(\tilde{\chi})$, included in the third panel of Figure S4 (a), constitutes a simulated FMR spectrum, with the absorption lines centred in the same $B_{RES}$ values determined through the SB formalism.

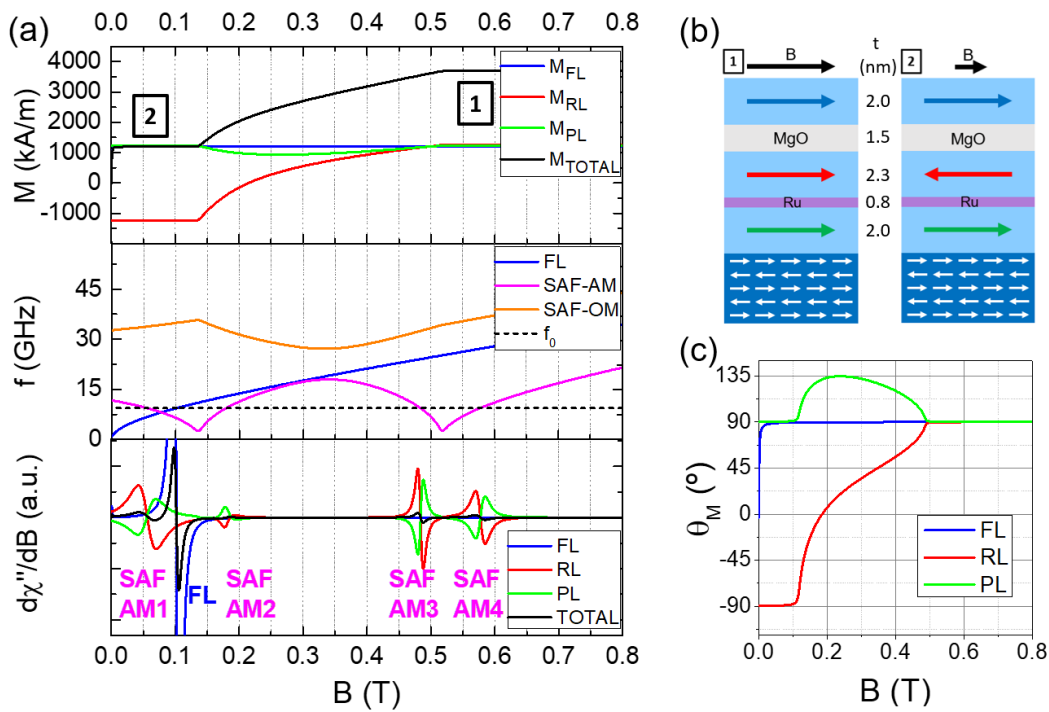

Figure S4. (a) Simulation of magnetostatic curve (top panel), ferromagnetic resonance modes (middle panel) and first derivative of the microwave absorption curve (bottom panel) for an MTJ with parameters $B_{K1eff}^{FL}$=-1 T, $B_{K1eff}^{RL}$=-1.6 T (extrapolation from $B_{K1eff}^{FL}(t_{FL})$), $M_{PL}$ = 1232 kA/m, $J_{EB}$ = 0.459 mJ/m², $J_{IEC}$ = − 0.797 mJ/m², $\alpha_{FL}$ = 0.02, $\alpha_{RL}$ = 0.04, $\alpha_{PL}$ = 0.04, g = 2.11 and f = 9.87 GHz. The MTJ multilayer stack is depicted in (b) with the external field **B** applied in plane along the exchange-bias direction ($\phi_B = 0$). (c) In-plane rotation ($\theta_M$) of the magnetization in the RL and PL with decreasing field.

## SM4. Magnetoresistance loops of pristine MTJ

The R(H) loops obtained by CIPT technique for pristine MTJs are shown in figure S5. The parallel (antiparallel) state with a low (high) resistance is observed at the positive (negative) values of H. $R_P$ and $R_{AP}$ of figure S5 (a) thus correspond, respectively, to the magnetic configurations *(2)* and *(3)* of figure S3 (a). The center of the R(H) loop is offset by approximately $-0.2$ mT, indicating a $J_{\text{Néel}} < 2 \times 10^{-4}$ mJ/m$^2$.

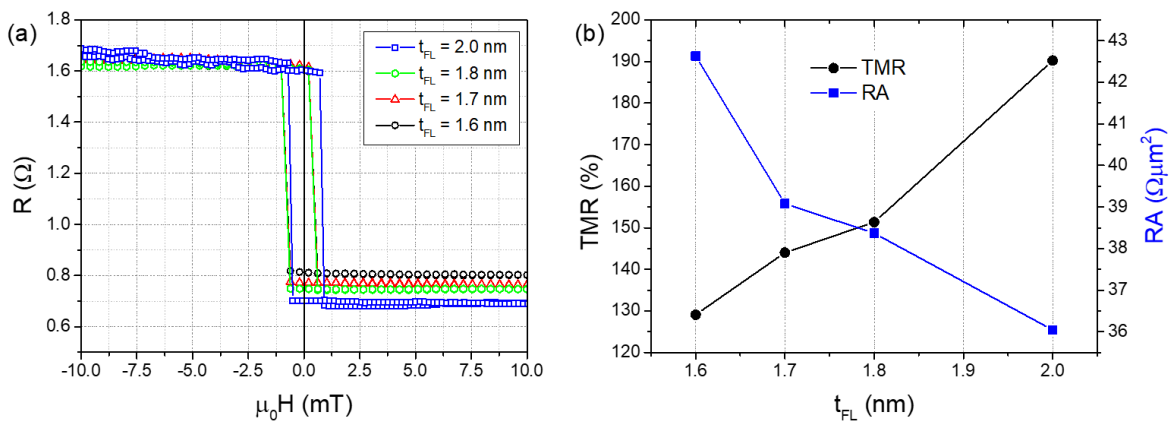

Figure S5. (a) R(H) loops for different thicknesses of the free layer and (b) TMR and RA as a function of the free layer thickness.

## SM5. Simulation of SAF-AM1 after irradiation at low fluences

The spectra simulated in figure S6 shows the shift of the absorption line corresponding to the SAF-AM1 mode with the decreasing exchange-bias coupling. The same tendency was observed in the experiment.

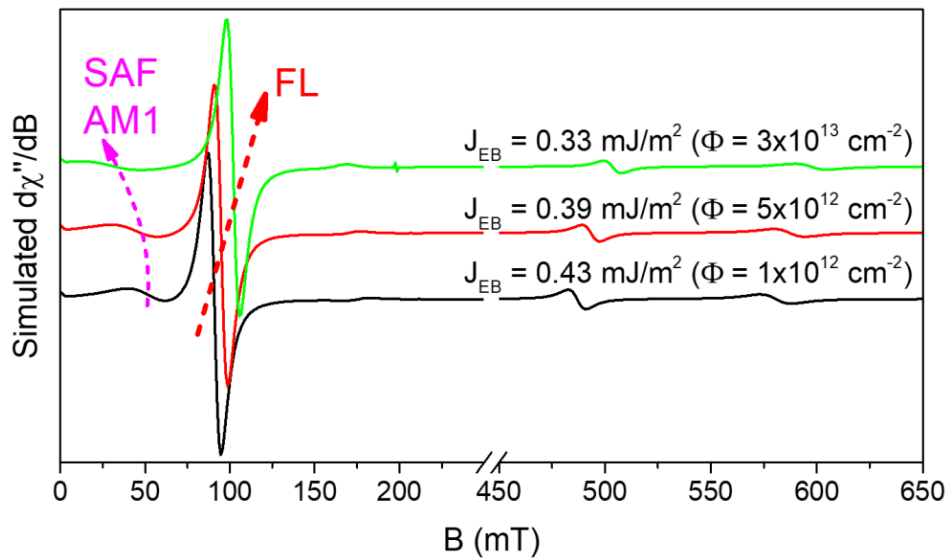

Figure S6. Simulated FMR absorption curves at $\phi_B = 0°$ for a $J_{EB}$ decreasing with the ion fluence according to experimental results. Below $\Phi = 10^{14}$ cm$^{-2}$, the SAF-AM1 mode shifts to lower field values with increasing $\Phi$, as seen also in the experiment. All other simulation parameters, apart from $B_{K1eff}^{FL}$ and $J_{EB}$, were kept constant.

Bruno Teixeira                                                                                                          09/04/2020

## SM6. X-Ray Reflectivity

The normalized XRR profile for the case of pristine and irradiated MTJ is presented in figure S7 (a). The position of the interference Kiessig fringes is practically unchanged upon ion irradiation, suggesting the thickness of the layers is not significantly changed. The irradiated MTJs exhibit, however, a faster decay of the whole reflectivity profile with the increasing angle as well as a decrease in intensity of the Kiessig fringes, both effects suggesting an increased interface roughness following the increasing ion fluence. We do not make a quantitative assessment of the XRR results, considering the difficulty in satisfactorily fitting all the fringes in the reflection profile. That difficulty mostly arises from the complexity of the multilayer stack, regarding the number of layer/interfaces, which only gets more complex after ion irradiation. The scattering length density, SLD, (scattering power of the material which increases with the electron density) profile of figure S7 (b) resulted from the best fits to the data of figure S7 (a). The changes in SLD suggest intermixing of the interfaces, although no clear conclusion on the tendency of the intermixing degree with the increasing ion fluence can be drawn.

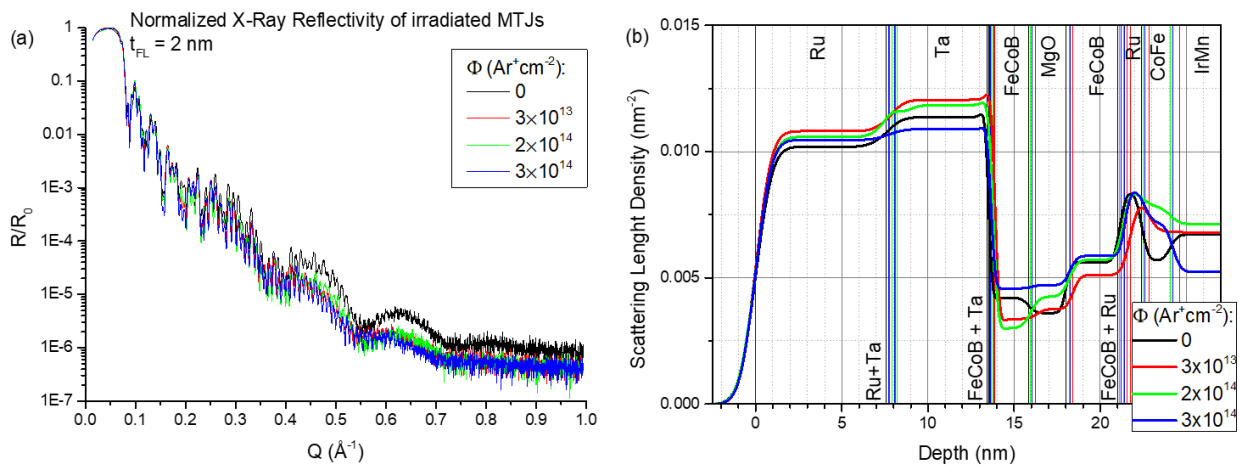

Figure S7. (a) Measured X-ray reflectivity profile and (b) calculated scattering length density for pristine MTJ (black) and Ar$^+$ irradiated MTJs with fluences $3 \times 10^{13}\ cm^{-2}$ (red), $2 \times 10^{14}\ cm^{-2}$ (green) and $3 \times 10^{14}\ cm^{-2}$ (blue).

## SM7. Role of mode hybridization on the linewidth asymmetry

The simulation presented in figure S8 shows how the peaks of microwave absorption corresponding to the SAF (RL and PL) are overlapped with the resonance of the FL, for the case of the AP state of the MTJ ($\phi_B = 0°$), but not for the P state ($\phi_B = 180°$). It is thus expected that any mode hybridization originated by $J_{Néel}$ may impact the FMR linewidth differently for each orientation.

The value of $J_{Néel}$ was increased in the simulations, and the linewidth was extracted by fitting Lorentzian lineshape to the simulated absorption curves. It is seen (figure S9 (a)) that for $J_{Néel} < 0.03 \text{ mJ/m}^2$, $\Delta B_{PP}$ increase less than 1%. Hence, the experimental estimate of $J_{Néel} \approx 2 \times 10^{-4} \text{ mJ/m}^2$ cannot explain the linewidth asymmetry seen for the pristine MTJ (figure 11(a) of the main text).

Afterwards, the evolution of the linewidth asymmetry with the increasing ion fluence was simulated (figure S9 (b)). $J_{Néel}$ was kept fixed at $2 \times 10^{-4} \text{ mJ/m}^2$, $2 \times 10^{-3} \text{ mJ/m}^2$ and $2 \times 10^{-2} \text{ mJ/m}^2$, while the magnetic anisotropy and the $J_{EB}$ and $J_{IEC}$ were varied according to the irradiation's experimental results. It was confirmed that $J_{Néel} = 2 \times 10^{-4} \text{ mJ/m}^2$ cannot explain the experimentally determined linewidth asymmetry and that a value as high as $J_{Néel} = 2 \times 10^{-2} \text{ mJ/m}^2$ would actually result in a decrease in $\Delta B_{PP}$ for $\phi_B = 0°$, up to a fluence of $10^{14} \text{ cm}^{-2}$, an opposite effect to that observed in the experiment.

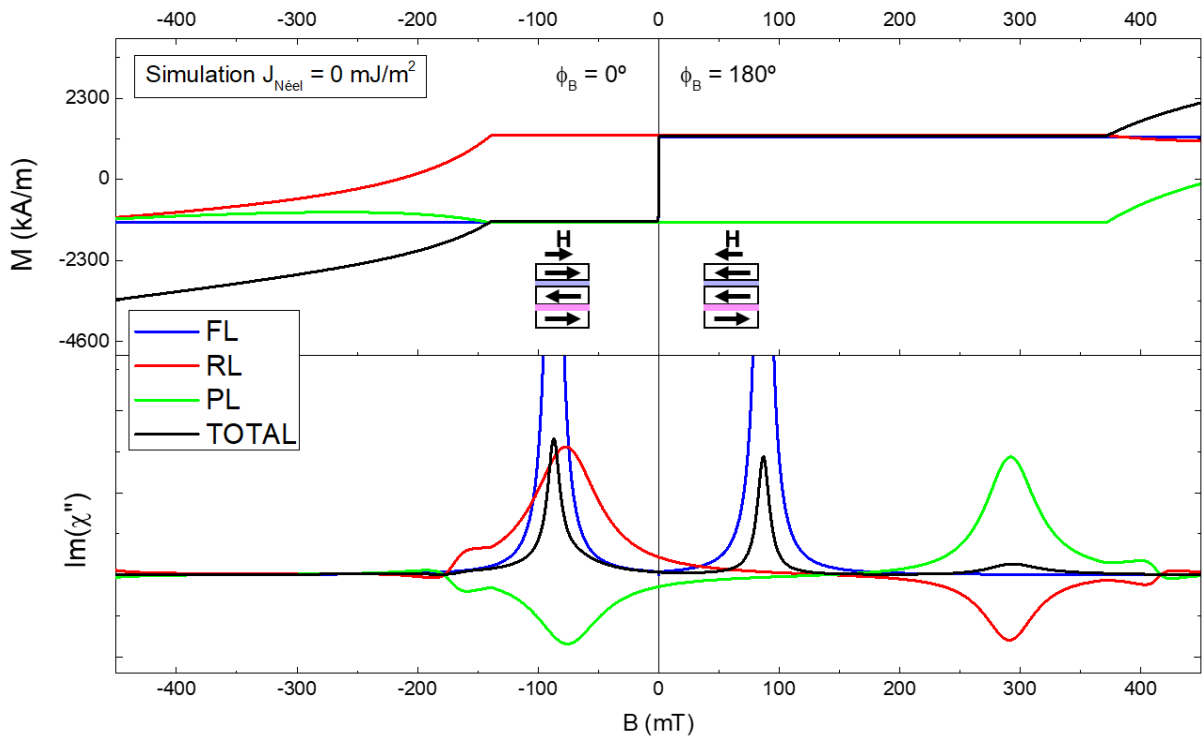

Figure S8. Simulated magnetization curve (top panel) and microwave absorption curves (bottom panel), for a pristine MTJ (simulation parameters as used in section SM3) and for a $J_{Néel} = 0 \ mJ/m^2$. In both cases, the total response of the MTJ is represented by the black line, while the individual responses of the FL, RL and PL are shown, respectively, in blue, red and green.

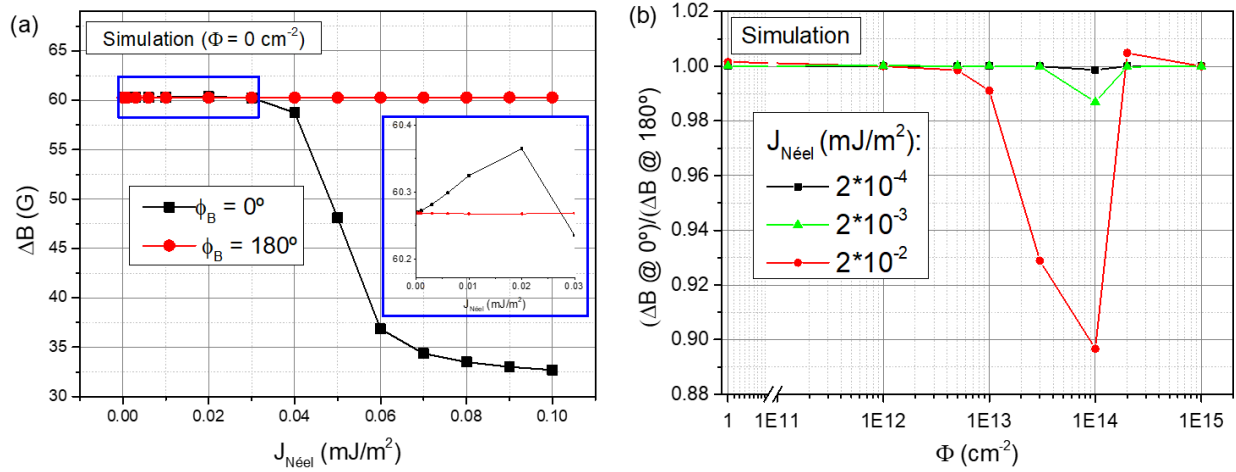

Figure S9. Simulations of: (a) FMR linewidth vs $J_{néel}$, for the AP (black squares) and P (red dots) configuration of the MTJ; and (b) linewidth asymmetry vs ion irradiation fluence for a $J_{néel} = 2 \times 10^{-4} mJ/m^2$ (black squares), $J_{néel} = 2 \times 10^{-3} mJ/m^2$ (green triangles), and $J_{néel} = 2 \times 10^{-2} mJ/m^2$ (red dots).